\documentclass{article}
\usepackage{amsmath,amssymb,amsthm}
\usepackage{graphicx}

\usepackage{url}
\usepackage{enumitem} 

\numberwithin{equation}{section}

\theoremstyle{plain}
\newtheorem{theorem}{Theorem}[section]
\newtheorem{lemma}[theorem]{Lemma}
\newtheorem{proposition}[theorem]{Proposition}
\newtheorem{corollary}[theorem]{Corollary}

\theoremstyle{definition}
\newtheorem{defn}[theorem]{Definition}

\newtheorem{example}[theorem]{Example}

\newcommand{\colim}{\mbox{\rm colim}}
\newcommand{\Gr}{\mbox{\rm Gr}}
\newcommand{\EA}{\mbox{\bf EA}}
\newcommand{\GEA}{\mbox{\bf GEA}}
\newcommand{\BCM}{\mbox{\bf BCM}}

\newcommand{\POG}{\mbox{{\bf POG}}}

\usepackage{tikz-cd}
\tikzset{%
    symbol/.style={%
        draw=none,
        every to/.append style={%
            edge node={node [sloped, allow upside down, auto=false]{$#1$}}}
    }
}
\tikzcdset{scale cd/.style={every label/.append style={scale=#1},
    cells={nodes={scale=#1}}}}

\author{Dominik Lachman\thanks{Department of Algebra and Geometry,\\
Palacký University Olomouc,\\
17. Listopadu 12,\\
Czech Republic,\\
\texttt{dominik.lachman@upol.cz}}%
\space\thanks{Author acknowledge the support by the Czech Science Foundation (GAČR): project 24-14386L}}

\title{Tensor Product in the Category of Effect Algebras}
\date{}

\begin{document}
\maketitle
\begin{abstract}
We study a tensor product in the category of effect algebras and in the category of partially ordered Abelian groups with order unit. We show that the tensor product preserves all the constructions that are essentially colimits over a connected diagram. Further, we prove the construction of a universal group for an effect algebra preserves all tensor products. We establish the corresponding functor from the category of effect algebras to the category of unital Abelian po-groups as a strong monoidal functor. We note that the technique we use in establishing the result could be used in various similar situations. Finally, we show that the tensor product of effect algebras does not preserve the Riesz decomposition property, which was an open question for a while.
\end{abstract}
\noindent \textbf{Mathematics Subject Classification:} 03G12, 06F20, 18M05.\\
\noindent \textbf{Keywords:} Effect algebras, Tensor product, Partially ordered Abelian groups, Riesz Decomposition Property, Monoidal categories.


\section{Introduction}\label{sec:intro}

In algebraic quantum logic, we associate certain ordered algebras to a physical system in such a way that elements of the algebra correspond to propositions about the system. In the seminal paper~\cite{BiNeu}, von Neumann and Birkhoff associate to a quantum mechanical physical system modelled on a Hilbert space $\mathcal{H}$ its logic $P(\mathcal{H})$. Elements of  $P(\mathcal{H})$ are projections to closed subspaces of $\mathcal{H}$. Then, they studied the algebraic structure that $P(\mathcal{H})$ naturally admits.

Since projections in $P(\mathcal{H})$ correspond to sharp propositions, we can consider so-called effects as unsharp propositions. In the situation modelled on a Hilbert space $\mathcal{H}$, the set of effects $E(\mathcal{H})$ consists of all self-adjoint operators between $0_\mathcal{H}$ and $1_\mathcal{H}$. Whereas the spectrum of a projection is a subset of $\{0, 1\}$,
the spectrum of an effect is a subset of the real interval $[0, 1]$. Effect algebras, introduced by Foulis and Bennett in~\cite{BF}, are an algebraic abstraction of the classical model $E(\mathcal{H})$, and they provide a general framework to study unsharp measurements in quantum mechanics. The class of effect algebras is quite vast since we can view many algebraic structures as effect algebras, in particular MV-algebras, orthoalgebras, or any interval $[0,u]$ of an Abelian partially ordered group $A$ (where $u\in A^+$). We also note that effect algebras gained attention in several recent interesting approaches to quantum logic. For example, they were recognized as certain Frobenius algebras in the category of relations, see \cite{Dus}, and they are essential to the effectus theory, see~\cite{effectus}.

In this article, we study a tensor product in the category $\EA$ of effect algebras. For quantum reasoning, the tensor product is an important concept. A tensor product of two Hilbert spaces embodies compounding the two corresponding physical systems together. In the algebraic approach to quantum logic, the concept of a tensor product was intensively studied as well. For example, in~\cite{FGR}, authors praise orthoalgebras as the smallest category which contains all orthomodular lattices and where each pair of orthomodular lattices has (in some sense) a convenient tensor product. In $\EA$, a tensor product of two effect algebras $E$ and $F$ is defined by its universal property: it is an effect algebra $E\otimes F$ such that each bimorphism based on $E\times F$ uniquely splits through $E\otimes F$. There are several articles concerning the tensor product in $\EA$ (e.g., \cite{JePu},\cite{Gu},\cite{JM}), yet it is still quite unclear how to compute a tensor product of two general effect algebras. Even the theorem which establishes the mere existence, proved in~\cite{Dv}, applies a nonconstructive argument.

        This article follows the opinion that some essential results concerning effect algebras and related structures are most effectively expressed in the language of category theory. For example, effect algebras are often studied using test spaces. An accurate description of the relation between these two concepts is that we have a reflection from the category of effect algebras to the category of test spaces. Similarly, the concept of the tensor product, which we define by a universal property, is categorical, and by proving the existence of tensor products we essentially equip a category with a monoidal structure.

When a symmetric monoidal category $\mathcal{C}$ admits a right adjoint to one-side tensoring $A\otimes-\colon\mathcal{C}\rightarrow\mathcal{C}$ given by the adjoint rule
\begin{align}\label{eq:MonHom}
 \begin{split}
       &A\otimes B\rightarrow C\\
        \hline
         &B\rightarrow [A,C]
    \end{split}
\end{align}
 we call it a closed monoidal category. The right adjoint is typically interpreted as an internal hom-functor.
In this case, tensoring is a left adjoint, so it preserves all the constructions that are essentially colimits. In $\EA$, the tensor product of effect algebras does not preserve ordinal sums (see Example~\ref{ex:cop}).
Since ordinal sums are coproducts, the category of effect algebras is not a closed monoidal category. As Jacobs and Mademaker point out in~\cite{JM}, one cannot give $\mathrm{Hom}(E, F)$ an effect algebra structure for arbitrary effect algebras $E$ and $F$ , which is in contrast to the case of partial commutative monoids.
Nevertheless, there are several results about colimits preserved by tensoring in $\EA$, e.g., see~\cite{JePu} for certain directed colimits. One of our main results provides a characterization of colimits preserved by tensor product and gives a new light on the situation. We show that 
$E\otimes-$ admits a right adjoint when we consider it as a functor from $\EA$ to an under category $E\downarrow\EA$, sending $F$ to $(E\rightarrow E\otimes F)$. Using 
this observation, we prove that $E\otimes-$ preserves all colimits over connected diagrams (in particular directed ones).

So far, thanks to various researchers, we have many results about various tensor products in several categories having a connection to $\EA$, that is, we have many monoidal categories around $\EA$: see~\cite{FB2} for orthoalgebras, \cite{Pu} for divisible effect algebras, \cite{JePu} for dimension effect algebras, \cite{Gu2} for sequential effect algebras, \cite{Gu} for monotone $\sigma$-complete effect algebras. In contrast, there are significantly fewer results about which functors between the involved categories (e.g., construction of universal group) well interact with tensor products, that is, which functors are monoidal. From a category theory point of view, functors and natural transformations are what really matter. In this sense, the situation, as described above, seems to be quite unsatisfactory and there is a lot to be done.

To give an example of the kind of results we seek, consider the tensor product in the category of partially ordered Abelian groups which we denote $\POG$. For $A,B\in\POG$ with positive cones $A^+$ and $B^+$, we have $(A,A^+)\otimes (B, B^+)\cong (A\otimes B,A^+\otimes B^+)$, that is, the forgetful functor $\POG\rightarrow \mathbf{Ab}$ is strong monoidal. This result essentially follows from a result by Goodearl and Handelman, Proposition 2.1., in~\cite{GH}. Some other results of this kind are given by Wehrung in~\cite{W} for categories of abelian po-groups and ordered commutative monoids. Another example of such a result is in~\cite{Pu}, where Pulmannová essentially proved that the construction of a universal group induces a strong monoidal functor between the categories of divisible effect algebras and Abelian po-groups.

Our next main result contributes to the list above by establishing the universal group construction as a strong monoidal functor $\Gr\colon\EA\rightarrow\POG_u$, where $\POG_u$ is the category of Abelian po-groups with an order unit. Moreover, the technique we use to establish this result has a strong potential to be useful in various similar situations.

 In the already mentioned article~\cite{W}, a lot of attention is paid to the problem of whether the tensor product of Abelian po-groups and partially ordered commutative monoids preserves the Riesz decomposition property (called interpolation in the context of po-groups and refinement property in the context of monoids). It turns out that the tensor product does preserve (RDP) in the case of monoids, while in the case of po-groups, there is a counter-example. The case of effect algebras was an open problem for a while, even if it was generally conjectured (RDP) is not preserved by tensoring (see~\cite{JePu}). By our result, the construction of the universal group $\Gr\colon \EA\rightarrow\POG_u$ extends to a strong monoidal functor. This provides a convenient bridge which enables us to lift Wehrung's counter-example in $\POG_u$ to $\EA$. This result indicates that the construction of tensor products of effect algebras is hard, in the sense that we cannot control it using (RDP). This is in contrast to the construction of a universal group which preserves (RDP). 

 \section{Preliminary}
In this section, we review some important concepts for this article. 
\begin{defn}
An effect algebra is a partial algebra $(E;\oplus,',0,1)$, where $\oplus$ is a partial binary operation, $'$ is a unary operation, and $0,1$ are constants such that for any $a,b,c\in E$:
\begin{enumerate}\item[(i)] $a\oplus b=b\oplus a$; \item[(ii)] $(a\oplus b)\oplus c=a\oplus(b\oplus c)$; \item[(iii)] $a\oplus b=1$ if and only if $b=a'$; \item[(iv)] $a\oplus 1$ is defined if and only if $a=0$, in this case $0\oplus 1=1$.
\end{enumerate}
Here (i--ii) are Kleene identities (i.e., if one side is defined, then the other is also defined and equality holds).
\end{defn}
Given elements $a$, $b$ of an effect algebra $E$, if $a\oplus b$ is defined we write $a\perp b$. A relation $\leq$ defined by prescription $a\leq b\Leftrightarrow (\exists c) a\oplus c = b$ is a partial order, in particular, the positivity property holds: $a\oplus b=0$ implies $a=b=0$. Moreover, $\oplus$ satisfies the cancellation property; hence we can define the operation of partial subtraction denoted $\ominus$. Summing up, effect algebras are precisely cancellative bounded partial commutative monoids satisfying the positivity property. If we leave the assumption of the top element, we obtain a concept of generalized effect algebras:
\begin{defn}
    A generalized effect algebra is a partial algebra $(E;\oplus,0)$, where $\oplus$ is a partial binary operation and $0$ is constant such that for any $a,b,c\in E$:
\begin{enumerate}
\item[(i)] $(E;\oplus,0)$ is a partial commutative monoid; 
\item[(ii)] $a\oplus b=a\oplus c\implies b=c$ (cancellation property); 
\item[(iii)] $a\oplus b=0\implies a=0$ (positivity property).
\end{enumerate}
\end{defn}
Note that given a generalized effect algebra $E$, any interval $[0, a]\subseteq E$ is naturally an effect algebra. One of our observations is that generalized effect algebras naturally arise when we study effect algebras.

By a \emph{momorphism of effect algebra} we mean a mapping between two effect algebras $f\colon E\rightarrow F$ such that: (1) for all $a,b\in E$, if $a\perp b$ then $f(a)\perp f(b)$, and $f(a\oplus b)=f(a)\oplus f(b)$ and (2) $f(1)=1$. Consequently, an effect algebra homomorphism preserves $0$ and $'$ as well. For the case of \emph{generalized effect algebra homomorphism}, we only assume condition (1). Let us denote $\EA$ the category of effect algebras with effect algebra homomorphisms. Similarly, we denote $\GEA$ the category of generalized effect algebras.

A useful consequence of the cancellation property is the following lemma:
\begin{lemma}\label{lem:vol}
Let $E$ be an effect algebra and $a,b,c,d\in E$ a quadruple of its elements such that $a\perp b$, $a\geq c$, and $b\geq d$. Then 
\begin{equation}\label{eq:vol}
    (a\oplus b)\ominus (c\oplus d)=(a\ominus c)\oplus (b\ominus d).
\end{equation}
\end{lemma}
\begin{proof}
     If we add $(c\oplus d)$ to both sides, we get in both cases $a\oplus b$. Hence, by the cancellation property~\eqref{eq:vol} holds.  
\end{proof}
Next, we recall the concept of a tensor product.
\begin{defn}
    Given three effect algebras $E,F$ and $G$, we call a mapping $b\colon E\times F\rightarrow G$ an effect algebra bihomomorphism if for each $a\in E$ and $b\in F$, we have
    \begin{enumerate}
        \item[(i)] $\beta(a,-)$ and $\beta(-,b)$ are homomorphisms of generalized effect algebras (preserve existing orthosums),
        \item[(ii)] $\beta(1,-)$ and $\beta(-,1)$ are homomorphisms of effect algebras ($b(1,1)=1$).
    \end{enumerate}
\end{defn}
\begin{defn}
    Given two effect algebras $E$ and $F$, their tensor product in $\EA$ is a bihomomorphism $(-\otimes-)\colon E\times F\rightarrow E\otimes F$ satisfying the following universal property: For any bihomomorphism $\beta\colon E\times F\rightarrow G$, there is a unique effect algebra homomorphism $h$ such that $\beta=h\circ\otimes$.
\end{defn}
The following theorem is proved in~\cite{DvPu}.
\begin{theorem}[\cite{DvPu}, Thm. 4.2.2]
    In $\EA$, a tensor product $E\otimes F$ exists for each pair of effect algebras $E$, $F$. Moreover, each element $c\in E\otimes F$ is an orthosum of simple tensors, that is
    \begin{equation*}
        c=a_1\otimes b_1\oplus\cdots \oplus a_n\otimes b_n,
    \end{equation*}
    for some $a_1,\ldots,a_n\in E$ and $b_1,\ldots,b_n\in F$.
\end{theorem}
Even if a tensor product is by definition a bihomomorphism, as usual, we will refer to the object part $E\otimes F$ as the tensor product of $E$ and $F$. In the next section, we will often use the fact that the tensor product behaves functorially. Let us describe the action of the tensor product on morphisms. Given two effect algebra homomorphisms $g\colon E_1\rightarrow E_2$ and $h\colon F_1\rightarrow F_2$, from the universal property, there is a unique $g\otimes h\colon E_1\otimes F_1\rightarrow E_2\otimes F_2$ fitting in: 
\begin{equation}\label{diag:bifunctors}
    \begin{tikzcd}
        {E_1\times F_2}\arrow[d,"\otimes"]\arrow[r,"g\times h"]&{E_2\times F_2}\arrow[d,"\otimes"]\\
        {E_1\otimes F_1}\arrow[r,"g\otimes h"]&{E_2\otimes F_2}
    \end{tikzcd}
\end{equation}
The uniqueness easily implies that the tensor product is functorial in both coordinates. Hence, for a fixed effect algebra $E$, we have an endofunctor $E\otimes -\colon \EA\rightarrow \EA$.

Note that the tensor product is defined up to isomorphism. Therefore, for example, the associativity rule holds up to isomorphism. This is well captured in the definition of a monoidal category, let us recall it.
\begin{defn}\label{def:monoidalCat}
By a monoidal category we mean a sextuplet $(\mathcal{C},\otimes,I,\alpha,\lambda,\rho)$ consisting of a category $\mathcal{C}$, a functor $\otimes\colon\mathcal{C}\times\mathcal{C}\rightarrow\mathcal{C}$, an object $I$ of $\mathcal{C}$, a natural  isomorphism $$\alpha\colon((-)\otimes(-))\otimes(-)\cong (-)\otimes((-)\otimes (-)),$$ a natural isomorphism $\lambda\colon I\otimes (-)\cong(-)$, and a natural isomorphism $\rho\colon (-)\otimes I\cong(-)$ so that for each $A$, $B$, $C$ and $D$ objects of $\mathcal{C}$, the following two diagrams commute:
     \begin{equation}\label{diag:triangle}
    \centering
\begin{tikzcd} 
{(A\otimes I)\otimes B} \arrow[rr,"\alpha_{A,I,B}"]\arrow[rd,"\rho_A\otimes B"]&&{A\otimes (I\otimes B)}\arrow[ld,"A\otimes\lambda_{B}"]\\ 
&{A\otimes B}&
 \end{tikzcd}
\end{equation}
     \begin{equation}\label{diag:pentagon}
    \centering
\begin{tikzcd} 
&{(A\otimes B)\otimes(C\otimes D)}\arrow[rd,"\alpha_{A.B,C\otimes D}"]&\\
{((A\otimes B)\otimes C)\otimes D}\arrow[d,"\alpha_{A,B,C}\otimes\mathrm{id}_D"]\arrow[ru,"\alpha_{A\otimes B,C,D}"]&&{A\otimes(B\otimes(C\otimes D))}\\ 
{(A\otimes (B\otimes C))\otimes D}\arrow[rr,"\alpha_{A,B\otimes C,D}"]&&{A\otimes((B\otimes C)\otimes D)}\arrow[u,"\mathrm{id}_A\otimes\alpha_{B,C,D}"]
 \end{tikzcd}
\end{equation}
\end{defn}
Monoidal categories important to this article satisfy an additional property that the monoidal unit and initial object coincide. These categories are called \emph{semi cocartesian monoidal categories}. 

\begin{defn}\label{def:iota}
    Assume a monoidal category where the monoidal unit $I$ is also the initial object. For any two objects $A$ and $B$ denote 
    \begin{equation}\label{eq:iota}
    \iota_{A,B}\colon A\rightarrow A\otimes B,
\end{equation}
which arise as the composition of $\rho^{-1}_A\colon A\rightarrow A\otimes I$ and $1_{A}\otimes !_B\colon A\otimes I\rightarrow A\otimes B$. 
\end{defn}
The following two examples of monoidal semi cocartesian categories are central to this article.
\begin{example}
    The category of effect algebras $\EA$ with the tensor product and $I$ equal to the two-element effect algebra $\underline{2}$  is a monoidal category. Moreover, Definition~\ref{def:iota} also applies to $\EA$, as $\underline{2}$ is an initial object. For $E,F\in\EA$, we have $\iota_{E,F}\colon a\mapsto a\otimes 1$.
\end{example}
By $\POG_u$ we denote the category of Abelian unital po-groups. That is, an object in $\POG_u$ is an Abelian po-group $A$ with a distinguished element $u\in A^+$ such that for each element $x\in A$, there is a natural number $n$, so that $x\leq n\cdot u$. We call $u$ satisfying this property an order unit, and we refer to the corresponding unital po-group as $(A,u)$. Morphisms in $\POG_u$ are positive group homomorphisms  which preserve order units.
\begin{example}
    The category of Abelian po-groups with an order unit $\POG_u$ forms a monoidal category, where $\otimes$ coincides with the standard tensor product of Abelian groups, $I$ equals $(\mathbb{Z},1)$, and ordering is induced by a cone $A^+\otimes B^+$. Again, $(\mathbb{Z},1)$ is also an initial object in $\POG_u$, and for $A=(A,u)$ and $B=(B,v)$ we have $\iota_{A,B}\colon a\mapsto a\otimes v$.
\end{example}

\begin{defn}\label{defn:monoidalFunctor}
    Let $(\mathcal{C},\otimes_{\mathcal{C}},I^\mathcal{C},\alpha^{\mathcal{C}},\lambda^\mathcal{C},\rho^{\mathcal{C}})$ and $(\mathcal{D},\otimes_{\mathcal{D}},I^\mathcal{D},\alpha^{\mathcal{D}},\lambda^\mathcal{D},\rho^{\mathcal{D}})$ be two monoidal categories. A lax monoidal functor from $\mathcal{C}$ to $\mathcal{D}$ is a functor $F\colon\mathcal{C}\rightarrow\mathcal{D}$ together with a morphism $\epsilon\colon I_D\rightarrow F(I_C)$ and a natural transformation $\mu\colon F(-)\otimes_\mathcal{D} F(-)\rightarrow F((-)\otimes_{\mathcal{C}}(-))$ so that the following three diagrams commute
    \begin{equation}\label{eq:unital}
    \centering
    \begin{tikzcd}[scale cd=0.96]
        {F(I^{\mathcal{C}}\otimes_{\mathcal{C}}A)}\arrow[d,"F(\lambda^{\mathcal{C}}_A)"]&
        {F(I^\mathcal{C})\otimes_{\mathcal{D}}F(A)}\arrow[l,"\mu_{I^\mathcal{C},A}"]&
        {F(A\otimes_{\mathcal{C}}I^{\mathcal{C}})}\arrow[d,"F(\lambda^{\mathcal{C}}_A)"]&
        {F(A)\otimes_{\mathcal{D}}F(I^\mathcal{C})}\arrow[l,"\mu_{A, I^\mathcal{C}}"]
        \\
        {F(A)}&{I^\mathcal{D}\otimes_{\mathcal{D}}F(A)}\arrow[u,"\epsilon\otimes\mathrm{id}_{F(A)}"]\arrow[l,"\lambda^\mathcal{D}_{F(A)}"]&{F(A)}&{F(A)\otimes_{\mathcal{D}}I^\mathcal{D}}\arrow[u,"\mathrm{id}_{F(A)}\otimes\epsilon"]\arrow[l,"\rho^\mathcal{D}_{F(A)}"]
    \end{tikzcd}
\end{equation}
\begin{equation}\label{eq:asso}
    \centering
    \begin{tikzcd}[scale cd=0.95]
        {(F(A)\otimes_\mathcal{D}F(B))\otimes_{\mathcal{D}}F(C)}\arrow[rrr,"\alpha^{\mathcal{D}}_{F(A),F(B),F(C)}"]\arrow[d,"\mu_{A,B}\otimes\mathrm{id}"]&&&{F(A)\otimes_\mathcal{D}(F(B)\otimes_\mathcal{D}F(D))}\arrow[d,"\mathrm{id}\otimes\mu_{C,D}"]\\
        {F(A\otimes_\mathcal{C}B)\otimes_{\mathcal{D}}F(C)}\arrow[d,"\mu_{A\otimes B,C}"]&&&{F(A)\otimes_\mathcal{D}F(B\otimes_\mathcal{C}D)}\arrow[d,"\mu_{A,B\otimes C}"]\\
        {F((A\otimes_\mathcal{C}B)\otimes_{\mathcal{C}}C)}\arrow[rrr,"\alpha^{\mathcal{C}}_{A,B,C}"]&&&{F(A\otimes_\mathcal{C}(B\otimes_\mathcal{C}D))}
    \end{tikzcd}
\end{equation}
If $\epsilon$ and $\mu$ are isomorphisms, we call $F$ a strong monoidal functor.
\end{defn}
We will need some additional categorical concepts which we now recall, namely \emph{under categories} (also called coslice categories) and \emph{property of finality}. Assume a category $\mathcal{C}$ and its object $A$. By $A\downarrow\mathcal{C}$ we denote a category which objects are morphisms in $\mathcal{C}$ having $A$ as the domain, and for two objects $f\colon A\rightarrow B$ and $g\colon A\rightarrow C$ a morphism from $f$ to $g$ is a morphism $h\colon B\rightarrow C$ from the original category $\mathcal{C}$ such that $h\circ f=g$. 

In a proof of one of our main theorems, we will also need a criterion of finality. Let us briefly introduce the concept. One can find more details as well as the proof of Lemma~\ref{lem:final} (in a more general situation) in~\cite{ARV}.
\begin{defn}
    We say that a subcategory $\mathcal{E}\subseteq\mathcal{D}$ is final (in $\mathcal{D}$) if for each functor $F\colon\mathcal{D}\rightarrow \mathcal{C}$ it holds: If $\colim_{d\in \mathcal{D}} F(d)$ exists, then $\colim_{d\in\mathcal{E}} F(d)$ exists as well, and they are (in a canonical way) isomorphic.
\end{defn}
We say that a category $\mathcal{D}$ is \emph{connected} if it is nonempty and connected as a graph, where we forget the orientation of arrows, that is, one can travel between any two objects following arrows but ignoring their orientation. 
\begin{lemma}\label{lem:final}
    Assume a small category $\mathcal{D}$ and its subcategory $\mathcal{E}\subseteq\mathcal{D}$. A category $\mathcal{E}$ is final if and only if for each $d\in\mathcal{D}$, the coslice category $d\downarrow \mathcal{E}$ is connected.
\end{lemma}

\section{Tensor product and connected colimits}
For two effect algebras $E$ and $F$, denote by $[E,F]$ the set of all generalized effect algebra homomorphisms from $E$ to $F$. That is, $[E,F]$ consists of mappings preserving all finite orthosums. On the set $[E,F]$ we can pointwise define relation $\perp$, partial operation $\oplus$, and partial order $\leq$ which yield a generalized effect algebra as the following lemma states.
\begin{lemma}
Let $E,F$ be effect algebras and $f,g\in [E,F]$. The following hold
\begin{enumerate}
    \item[(i)] if $f\perp g$, then the point-wise orthosum $f\oplus g$ belongs to $[E,F]$;
    \item[(ii)] if $f\leq g$, then then the point-wise subtraction $f\ominus g$ belongs to $[E,F]$.
\end{enumerate}
Consequently, $([E,F],\oplus,0_E)$ is a generalized effect algebra whose (algebraic) ordering coincides with the pointwise one.
\end{lemma}
\begin{proof}
For (i), first note that $f\oplus g$ is a well-defined mapping, so we only need to check it preserves finite orthosums. Mapping $f\oplus g$ clearly preserves $0$, and for any $a,b\in E$, $a\perp b$, we have
\begin{align*}
    (f\oplus g)(a\oplus b)&=f(a\oplus b)\oplus g(a\oplus b)=f(a)\oplus f(b)\oplus g(a)\oplus g(b)\\
    &=(f\oplus g)(a)\oplus (f\oplus g)(b).
\end{align*}

Next consider (ii), if $f\leq g$, then $f\ominus g$ is a point-wise well-defined mapping.
Clearly $(f\ominus g)(0)=0$, 
and it preserves orthosum as well, since for each $a,b\in E$, $a\perp b$, it holds
\begin{align*}
    &(g\ominus f)(a\oplus b)=g(a\oplus b)\ominus f(a\oplus b)=(g(a)\oplus g(b))\ominus (f(a)\oplus f(b))=\\
    = &(g(a)\ominus f(a))\oplus (g(b)\ominus f(b))    = (g\ominus f)(a)\oplus (g\ominus f)(b),
\end{align*}
where in the third equality we apply Lemma~\ref{lem:vol}.

Now, the statement (i) implies that $[E,F]$ endowed with $\oplus$ is a partial commutative monoid. Furthermore, $[E,F]$ obviously satisfies the cancellation property and positivity property; therefore, it is a generalized effect algebra. Statement (ii) asserts that $f\leq g$ point-wise iff $f\leq g$ as elements of a generalized effect algebra.
\end{proof}

\begin{lemma}\label{lem:intIsEA}
    Let $E,F$ be effect algebras and $h\in[E,F]$. Then the interval $[E,F]_h:=\{g\in[E,F] \mid  g\leq h\}$ is an effect algebra (with the obvious structure).
\end{lemma}
\begin{proof}
    The interval in concern is a generalized effect algebra with a top element $h$, hence it is an effect algebra.
\end{proof}
We are interested in an endofunctor $E\otimes -\colon\EA\rightarrow\EA$ for a fixed effect algebra $E$. In particular, which colimits it preserves. One can prove it does not preserve coproducts, hence it does not admit a right adjoint. In this sense, it is more convenient to consider it as a functor $\EA\rightarrow E\downarrow\EA$, where the codomain category is the category of arrows having $E$ as the domain. Tensoring with $E$ obviously determines a functor 
\begin{equation}\label{eq:loc0}\underline{2}\downarrow\EA \rightarrow E\otimes\underline{2}\downarrow\EA,\end{equation} which sends a morphism $\underline{2}\rightarrow F$ to $E\otimes\underline{2}\rightarrow E\otimes F$. As $\underline{2}$ is both the initial object of $\EA$ and the unit for the monoidal structure, we have the obvious isomorphisms $\EA\cong \underline{2}\downarrow\EA$ and $E\otimes\underline{2}\downarrow\EA\cong E\downarrow\EA$. Hence~\eqref{eq:loc0} is essentially
\begin{equation}\label{eq:loc1}\EA\rightarrow E\downarrow \EA.\end{equation}
Recalling~\eqref{eq:iota}, the functor~\eqref{eq:loc1} acts as $F\mapsto(\iota_{E,F}\colon E\rightarrow E\otimes F)$.
Using this arrangement the right adjoint exists:
\begin{theorem}\label{thm:adj}
For an effect algebra $E$,  the functor $E\otimes -$ from $\EA$ to $E\downarrow \EA$ which sends $F$ to a morphism $\iota_{E,F}\colon E\rightarrow E\otimes F$ ($a\mapsto a\otimes 1$) admits a right adjoint $[E,-]_-$ which sends an object $h\colon E\rightarrow G$ in $E\downarrow \EA$ to an effect algebra $[E,G]_h$ (established in Lemma~\ref{lem:intIsEA}).
\end{theorem}
\begin{proof}
A morphism $f\colon(E\rightarrow E\otimes F)\rightarrow (h\colon E \rightarrow G)$ in $E\downarrow \EA$ corresponds (bijectively) to a bimorphism $\beta\colon E\times F\rightarrow G$ which sends $(a, 1)$ to $h(a)$. We show that this bimorphism corresponds (bijectively) to an effect algebra homomorphism 
$$\overline{\beta}\colon F\rightarrow [E,G]_h$$
given by prescription $\overline{\beta}(b)\colon a\mapsto \beta(a,b)$. It is well known that such a prescription defines a natural bijection on the level of set mappings. Hence, we only need to verify that the bijection restricts to a bijection between the bimorphisms on one side and the homomorphisms on the other side.

Assume first that $\beta$ is a bimorphism, and let us prove $\overline{\beta}$ is a homomorphism. As $\beta$ preserves $\oplus$ in each coordinate, we have $\overline{\beta}(b)=\beta(-,b)$ belonging to $[E,G]_h$, for each $b\in F$. Next, given $b_1,b_2\in F$, $b_1\perp b_2$, 
we have $$\overline{\beta}(b_1\oplus b_2)=\beta(-,b_1\oplus b_2)=\beta(-,b_1)\oplus \beta(-,b_2)=\overline{\beta}(b_1)\oplus\overline{\beta}(b_2).$$ 
That is $\overline{\beta}$ preserves $\oplus$. Finally, we prove $\overline{\beta}$ preserves the bottom and the top elements $0$ and $1$. Mapping $\overline{\beta}(0)$ sends any $a\in E$ to $\beta(a,0)=0$, so $\overline{\beta}(0)$ is the bottom element in $[E,G]_h$. Further, $\overline{\beta}(1)(a)=\beta(a,1)=h(a)$. We deduce $\overline{\beta}(1)=h$, that is, $\overline{\beta}$ preserves the top element. We conclude $\overline{\beta}$ is an effect algebra homomorphism. 

To establish the converse direction, assume an effect algebra homomorphism $\overline{\beta}\colon F\rightarrow [E,G]_h$. We will check that a mapping $\beta\colon E\times F\rightarrow G$, given by $\beta(a,b)=\overline{\beta}(b)(a)$, is a bimorphism and $\beta(a,1)=h(a)$. For each $a_1,a_2\in E$, $a_1\perp a_2$, and $b\in F$, we have
$$\beta(a_1\oplus a_2,b)=\overline{\beta}(b)(a_1\oplus a_2)=\overline{\beta}(b)(a_1)\oplus\overline{\beta}(b)(a_2)=\beta(a_1,b)\oplus\beta(a_2,b).$$
Similarly, for each $b_1,b_2\in F$, $b_1\perp b_2$, and $a\in E$, we have
$$\beta(a,\beta_1\oplus \beta_2)=\overline{\beta}(b_1\oplus b_2)(a)=\overline{\beta}(b_1)(a)\oplus \overline{\beta}(b_2)(a)=\beta(a,b_1)\oplus\beta(a,b_2).$$
Finally, $\beta(1,1)=\overline{\beta}(1)(1)=h(1)=1$.
\end{proof}
Recall that the category $\EA$ is co-complete (see~\cite{JM}), and consequently, so is the under category $E\downarrow \EA$ for each effect algebra $E$. The following theorem characterises colimits which are preserved under tensoring. Let us state precisely the meaning of \emph{preserving a colimit} first. Assume a functor $\Phi\colon\mathcal{C}\rightarrow\mathcal{C}'$ and a small category $\mathcal{D}$. We say $\Phi$ preserves colimits over $\mathcal{D}$ if given any functor $F\colon\mathcal{D}\rightarrow \mathcal{C}$ and a colimiting cocone $(h_d\colon F(d)\rightarrow G)_{d\in\mathcal{D}}$ in $\mathcal{C}$, the cocone $(\Phi(h_d)\colon\Phi\circ F(d)\rightarrow \Phi(G))$ is a colimiting one as well. Equivalently, the induced morphism
\begin{equation}\label{eq:cononical}
    h\colon\colim_{d\in\mathcal{D}}\Phi\circ F(d)\rightarrow \Phi(\colim_{d\in\mathcal{D}}F(d))
\end{equation}
is an isomorphism.
It is well-known that $\Phi $ preserves all colimits whenever it admits a right adjoint.
\begin{example}\label{ex:cop}
    Let $E$ be a general effect algebra. Since $\underline{2}$ is both the initial object and the monoidal unit, we have
    $$E\otimes (\underline{2}\coprod\underline{2})\cong E\otimes\underline{2}\cong E\not\cong E\coprod E\cong E\otimes\underline{2}\coprod E\otimes\underline{2}.$$
    Consequently, tensoring by a general $E$ does not preserve coproducts.
\end{example}
\begin{theorem}\label{thm:conCol}
    Let $\mathcal{D}$ be a small category and $E\in\EA$. The functor $E\otimes-\colon \EA\rightarrow \EA$ preserves all colimits over $\mathcal{D}$ if and only if $\mathcal{D}$ is connected, or $E\cong \underline{2}$, or $E\cong \underline{1}$ and $\mathcal{D}$ is nonempty (where $\underline{1}$ is the one-element effect algebra, which is a terminal object in $\EA$).
\end{theorem}
\begin{proof}
First assume $\mathcal{D}$ is connected and let us prove $E\otimes-$ preserves colimits over $\mathcal{D}$.
The functor $E\otimes-$ admits an obvious splitting through~\eqref{eq:loc1} as 
\begin{equation}\label{eq:splitH}
    \EA \rightarrow E\downarrow\EA\rightarrow \EA,
\end{equation}
where the second functor is just the projection $(E\rightarrow F)\mapsto F$.
 The first functor in~\eqref{eq:splitH} is a left adjoint by Theorem~\ref{thm:adj}, so it preserves all colimits.   
 Hence, it is enough to prove the projection $E\downarrow \EA\rightarrow \EA$ preserves colimits over $\mathcal{D}$. 
 
 Assume a functor $F\colon \mathcal{D}\rightarrow E\downarrow\EA$. Define a category $\mathcal{D}^\bullet$ which arises from $\mathcal{D}$ by adding a new initial object $d_0$. The functor $F$ induces a functor $F^\bullet\colon \mathcal{D}^\bullet\rightarrow \EA$ as follows: On $\mathcal{D}$ we just compose $F$ with the projection $E\downarrow\EA\rightarrow \EA$, for $d_0$ we set $F^\bullet(d_0)=E$, and for each arrow $f\colon d_0\rightarrow d$ we define $F^\bullet(f)=F(d)$. Now, one can think through that the object part of $\colim_{d\in\mathcal{D}}F(d)$ (in $E\downarrow \EA$) arises by computing $\colim_{d\in\mathcal{D}^\bullet}F^\bullet(d)$ in $\EA$. Hence, the projection $E\downarrow\EA\rightarrow\EA$ preserves colimits over $\mathcal{D}$, whenever  $\mathcal{D}$ is a final subcategory of $\mathcal{D}^\bullet$. We will verify the condition in Lemma~\ref{lem:final}, i.e., for each $d\in\mathcal{D}^\bullet$, we show the category $d\downarrow\mathcal{D}$ is connected. For $d=d_0$, this easily follows from the assumption that $\mathcal{D}$ is connected. For $d\in\mathcal{D}$, the category $d\downarrow \mathcal{D}$ has an initial object $\mathrm{id}_d$, hence it is connected.

In the case $E\cong \underline{2}$, $\underline{2}\otimes -$ is isomorphic to the identity on $\EA$, hence it trivially preserves all colimits. Finally, let $E\cong\underline{1}$. We claim that for each $G\in\EA$, $\underline{1}\otimes G\cong\underline{1}$. Indeed, each homomorphism $f\colon \underline{1}\rightarrow G$, $G\in\EA$, is necessarily an isomorphism. Thus, so is the homomorphism $\iota_{\underline{1},G}\colon\underline{1}\cong\underline{1}\otimes{G}$. Consequently, $$\colim_{d\in\mathcal{D}}\underline{1}\otimes F(d)\cong \underline{1}\cong \underline{1}\otimes\colim_{d\in\mathcal{D}}F(d),$$ 
whenever $\mathcal{D}$ is non-empty.

 Next, assume $E\otimes-$ preserves all colimits over $\mathcal{D}$ and denote $\mathcal{D}=\coprod_{i\in I}\mathcal{D}_i$ the decomposition of $\mathcal{D}$ into its connected components. Set $F\colon\mathcal{D}\rightarrow \EA$ to be constantly equal $\underline{2}$. Since $\underline{2}$ is the initial object, the obvious cocone $(\mathrm{id}:\underline{2}\rightarrow \underline{2})_{d\in\mathcal{D}}$ is a colimiting one. When applying $E\otimes-$, we essentially obtain a cocone 
$(\mathrm{id}\colon E\rightarrow E)_{d\in\mathcal{D}}$, as $\underline{2}$ is the monoidal unit. However, the colimit of $E\otimes F(-)$ is 
$$\colim_{d\in\mathcal{D}}E\otimes F(d)= \colim_{d\in\mathcal{D}}E\otimes \underline{2}\cong\colim_{d\in\mathcal{D}}E\cong \coprod_{i\in I}\colim_{d\in\mathcal{D}_i}E\cong \coprod_{i\in I} E.$$
 Therefore, the homomorphism \eqref{eq:cononical} becomes in this case the folding homomorphism $h\colon \coprod_{i\in I}E\rightarrow E$. Clearly, $h$ is an  isomorphism only if $|I|=1$, in which case $\mathcal{D}$ is connected, or $|I|=0$ and $E\cong \underline{2}$, or $|I|\geq 2$ and $E$ has no element $a\not=0,1$, that is, $E\cong\underline{2}$ or $E\cong\underline{1}$ (note that in $\EA$, a coproduct is a horizontal sum).
\end{proof}
In particular, Theorem~\ref{thm:conCol} assures tensor product in $\EA$ preserves all coequalizers. On the other hand, coproduct is an essential example of a colimit that is (in general) not preserved by tensor product in $\EA$. It follows from the proof of Theorem~\ref{thm:conCol} that given a triple of effect algebras $E,F_1,F_2$, the effect algebra $E\otimes (F_1\coprod F_2)$ is isomorphic to the colimit of the span: $$E\otimes F_1\xleftarrow{\iota_{E,F_1}} E\xrightarrow{\iota_{E,F_2}} E\otimes F_2.$$

\section{Unital Abelian po-groups}\label{sec:4}
An analogous adjunction, as described in Theorem~\ref{thm:adj}, arises (based on the author's as yet unpublished computations) in several other categories related to $\EA$. In this section, we investigate the category of unital partially ordered Abelian groups, $\POG_u$.

The relation between $\EA$ and $\POG_u$ is provided by the well-known adjunction (see~\cite{DvPu}):
\begin{equation}\label{diag:adj1}
    \centering
\begin{tikzcd} {\EA} \arrow[rr,bend right=15,"\Gr"'{name=R}]&&{\POG_u} \arrow[ll,bend right=15,"\Gamma"'{name=L}]\arrow[from=R, to=L, symbol=\dashv] \end{tikzcd}
\end{equation}
where $\Gamma$ is the so-called Mundici functor which sends $(G,u)$ to the interval effect algebra $\Gamma(G,u)$. While $\Gr$ is a functor (sometimes called Grothendieck functor) which associates to each effect algebra $E$ its universal group. 
Moreover, the unit of the adjunction is an epimorphism:
\begin{lemma}\label{lem:univ}
    For an effect algebra $E$, each element of $\Gamma(\Gr(E))$ is an orthosum of elements in the image of the unit $\eta_E$. Moreover, the elements in $\mathrm{Im}(\eta_E)$, considered as elements of $\Gr(E)$, generate $\Gr(E)$ as well.
\end{lemma}
\begin{proof}
    It is implicit from the proof of the existence of a universal group in~\cite{DvPu} (Theorem 1.4.12).
\end{proof}
The following well-known theorem is essentially proved in~\cite{Rav}.
\begin{theorem}\label{thm:RDP}
    The adjunction~\eqref{diag:adj1} restricts to an equivalence between the category of effect algebras with (RDP) and the category of unital Abelian po-groups with interpolation.
\end{theorem}

Let us remark that the functor $\Gr$ is neither full nor faithful. However, it splits through the category of so-called barred commutative monoids $\mathbf{BCM}$ (introduced in~\cite{JM}), so that the factor $i\colon \EA\hookrightarrow {\bf BCM}$ is full and faithful. In fact, the whole adjunction~\eqref{diag:adj1} splits through $\BCM$, so $i$ is a reflection. For more details, see the mentioned article~\cite{JM}.

\begin{defn}
    Let $(A,u),(B,v)$ be two unital Abelian po-groups. Denote $[A,B]$ the Abelian group of homomorphisms $f\colon A\rightarrow B$ endowed with a partial order induced by a cone $[A,B]^+$ of all the homomorphisms preserving the partial order.  Moreover, given a homomorphism of unital po-groups $h\colon (A,u)\rightarrow (B,v)$, we denote by $[A,B]_h$ the maximal subgroup of $[A,B]$ which admits $h$ as an order unit, i.e., $[A,B]_h$ contains those homomorphisms $f\in[A,B]$ which satisfy $f\leq n\cdot h$, for some $n\in\mathbb{N}$. 
\end{defn}
\begin{defn}\label{defn:biHomuG}
    A bihomorphism of unital Abelian po-groups is a mapping $\beta\colon (A,u)\times (B,v)\rightarrow (C,w)$ such that for each $a\in A$ and $b\in B$:
    \begin{enumerate}
        \item[(i)] $\beta(a,-)\colon B\rightarrow C$ and $\beta(-,b)\colon A\rightarrow C$ are group homomorphisms.
        \item[(ii)] If $a\in A^+$, $b\in B^+$, then $\beta(a,-)$ and $\beta(-,b)$ are po-group homomorphisms (hence $\beta(A^+,B^+)\subseteq C^+$).
        \item[(iii)] $\beta(u,-)$ and $\beta(-,v)$ are unital group homomorphisms ($\beta(u,v)=w$).
    \end{enumerate}
\end{defn}
\begin{defn}
    Given two unital Abelian po-groups $(A,u)$ and $(B,v)$ in $\POG_u$, their tensor product in $\POG_u$ is a bihomomorphism $(-\otimes-)\colon (A,u)\times(B,v)\rightarrow (A\otimes B, u\otimes u)$ satisfying the following universal property: For any bihomomorphism $\beta\colon (A,u)\times(B,v)\rightarrow (C, w)$, there is a unique unital po-group homomorphism $h$ such that $\beta=h\circ\otimes$.
\end{defn}
The following theorem is implicit from~\cite{W}.
\begin{theorem}\label{thm:tensorPOG}
    In $\POG_u$, each pair of objects $(A,u)$ and $(B,v)$ admits a tensor product, which makes $\POG_u$ a monoidal category with unit $(\mathbb{Z},1)$. Moreover, the tensor product of two po-groups $(A,u),(B,v)$ is isomorphic to $(A\otimes B,u\otimes v)$, where $A\otimes B$ is a tensor product in the category of Abelian 
    groups (without partial order), with partial order given by a cone of all tensors $a_1\otimes b_1+\cdots+a_n\otimes b_n$, where for each $i=1,\ldots, n$, 
    $a_i\in A^+$ and $b_i\in B^+$.
\end{theorem}
Similarly as with $\EA$, we can use the semi cocartesian structure of $\POG_u$ to define for a given $A=(A,u)\in\POG_u$ a functor $(A,u)\otimes -\colon \POG_u\rightarrow (A,u)\downarrow \POG_u$, which sends $(B,v)$ to $\iota_{A,B}\colon (A,u)\rightarrow (A,u)\otimes(B,v)$. The following theorem is an analogue of Theorem~\ref{thm:adj}.
\begin{theorem}
    Let $A=(A,u)$ be a unital Abelian po-group and consider $(A,u)\otimes -$ as a functor from $\POG_u$ to $A\downarrow \POG_u$. Then $(A,u)\otimes -$ admits a right adjoint $[A,-]_{-}$ which sends $h\colon A\rightarrow B$ to $[A,B]_h$.
\end{theorem}
\begin{proof}
    Similarly as in the proof of Theorem~\ref{thm:adj}, we prove that the obvious bijection between sets of mappings $\{\beta\colon  A\times B\rightarrow C\mid \beta(a,v)=h(a)\}$ and  $\{g\colon B\rightarrow C^A \mid g(v)=h\}$ restricts to a bijection between the convenient bi-homomorphisms and homomorphisms.

    First, assume a bi-homomorphism $\beta\colon A\times B\rightarrow C$. The conditions (i--iii) on $\beta$ from Definition~\ref{defn:biHomuG} directly translate to $b\mapsto \beta(-,b)$ being a homomorphism of unital Abelian po-groups. We only need to check that $\beta(-,b)$ is dominated by some $n\cdot h$, $n\in\mathbb{N}$. But that is easy, take $n$ great enough so that $b\leq n\cdot v$. Then $c:=n\cdot v-b$ belongs to $B^+$ and so $\beta(-,c)$ is a positive element of $[A,C]_h$ such that $\beta(-,b)+\beta(-,c)=n\cdot \beta(-,v)$.

    Conversely, assume a unital po-group homomorphism $f\colon B\rightarrow [A,C]_h$ which yields a mapping $\beta\colon A\times B\rightarrow C$ defined as $\beta(a,b)=f(b)(a)$. We prove $\beta$ satisfies (i--iii) of Definition~\ref{defn:biHomuG}. For (i), $\beta$ is obviously additive in both coordinates, and for each $a\in A$, we have $\beta(a,0)=0$, since $\beta(-,0)=f(0)$ is the trivial homomorphism. Similarly, for each $b\in B$, there is $\beta(0,b)=0$ since $\beta(-,b)=f(b)$ is a homomorphism. Consider (ii), for any $a\in A^+$ and $b\in B^+$, we have $f(b)$ preserving ordering, and so $\beta(a,b)=f(b)(a)\in C^+$. Finally, (iii) follows from $\beta(u,v)=f(v)(u)=h(u)=w$.
\end{proof}
\begin{corollary}
    In  $\POG_u$, tensor product preserves all connected colimits.
\end{corollary}
\begin{proof}
    The proof is the same as that of Theorem~\ref{thm:conCol}.
\end{proof}

Denote $\POG_p$ the category where an object is an Abelian po-groups with distinguished positive element  $(A,x)$ (not necessarily strong unit) and a morphism $h\colon (A,x)\rightarrow (B,y)$ is a po-group homomorphism such that $h(x)=y$. Note that the resulting category is isomorphic to $\mathbb{Z}\downarrow\POG$. The obvious inclusion $i\colon\POG_u\hookrightarrow \POG_p$ has a right adjoint $R\colon\POG_p\rightarrow \POG_u$ given by $R\colon(A,x)\mapsto(A_0,x)$, where $A_0\leq A$ is the maximal subgroup so that $x$ is a strong unit for it. If we compose the adjunction $\Gamma\dashv\Gr$ with the just described adjunction, we obtain a natural isomorphism
\begin{equation}\label{eq:adjPOGp}
    \mathrm{Hom}_{\POG_p}((\Gr(E),1),(A,x))\cong\mathrm{Hom}_{\EA}(E,\Gamma(A,x)),
\end{equation}
for any Abelian po-group $A$ and $x\in A^+$.

For two unital Abelian po-groups $A=(A,u)$ and $B=(B,v)$, we denote $[A,B]_{\mathrm{GEA}}$ the set of po-group homomorphisms $g\colon A\rightarrow B$ with $g(u)\leq v$. We can naturally endow $[A,B]_{\mathrm{GEA}}$ with a structure of generalized effect algebra, where $g\perp h$ iff $g(u)+ h(u)\leq v$ and all operations are defined pointwise. One can read the following lemma as an internalisation of the adjunction $\mathrm{Gr}\dashv\Gamma$.
\begin{lemma}
    Let $E$ be an effect algebra and $(B,v)$ a unital Abelian po-group. Then there is an isomorphism of generalized effect algebras
    \begin{equation}\label{eq:internalAdj}
        [E,\Gamma(B,v)]\cong [\mathrm{Gr}(E),(B,v)]_{\mathrm{GEA}}
    \end{equation}
    which is natural in both $E$ and $(B,v)$.
\end{lemma}
\begin{proof}
    The set $[E,\Gamma(B,v)]$ decomposes as a disjoint union 
    $$\bigcup_{0\leq x\leq v}\mathrm{Hom}(E,\Gamma(B,x)).$$ Similarly, the right-hand side decomposes as  
    $$[\mathrm{Gr}(E),(B,v)]_{\mathrm{GEA}}=\bigcup_{0\leq x\leq v}\mathrm{Hom}(\mathrm{Gr}(E),(B,x)).$$ 
    For each fixed $x\in B^+$, $0\leq x\leq v$, the corresponding sets of homomorphisms are in the bijection provided by~\eqref{eq:adjPOGp}. Moreover, by the naturality of~\eqref{eq:adjPOGp}, the bijection~\eqref{eq:internalAdj} in concern is natural in both $E$ and $(B,v)$. 
    
    It remains to prove that the bijection induces an isomorphism of generalized effect algebras. The right-to-left direction proceeds as follows: 
\begin{equation}\label{eq:unit}
\Phi\colon  g\mapsto \Gamma(g)\circ \eta_E,    
\end{equation}
where $\eta_E\colon E\rightarrow \Gamma(\Gr(E),1)$ is the unit of the adjunction $\Gr\dashv\Gamma$. As $\Gamma(g)$ is just a restriction of $g$, we will abbreviate~\eqref{eq:unit} as a composition $\Phi\colon g\mapsto g\circ\eta_E$. We already know that $\Phi$ is a bijection and it obviously preserves the zero element. Moreover, it preserves and reflects the orthogonality relation: For each $g_1,g_2\in [\mathrm{Gr}(E),(B,v)]_{\mathrm{GEA}}$ we have ($u$ denotes the order unit in $\Gr(E)$)
\begin{align*}
    g_1\perp g_2&\Longleftrightarrow g_1(u)+g_2(u)\leq v\Longleftrightarrow  g_1(\eta_E(1))+g_2(\eta_E(1))\leq v\\
    &\Longleftrightarrow\Phi(g_1)(1)\perp\Phi(g_2)(1)\Longleftrightarrow\Phi(g_1)\perp\Phi(g_2).
\end{align*}
Finally, we show $\Phi$ preserves orthosums. For each pair $g_1, g_2\in [E,\Gamma(B,v)]$, $g_1\perp g_2$, we have
$$\Phi(g_1\oplus g_2)=(g_1\oplus g_2)\circ\eta_E=g_1\circ\eta_E\oplus g_2\circ\eta_E=\Phi(g_1)\oplus\Phi(g_2).$$
\end{proof}

\section{Tensor product in $\EA$ does not preserve (RDP)}
For a given effect algebra $E$ and a po-group $A=\mathrm{Gr}(E)$, the adjunction between $\EA$ and $\POG_u$ induces an adjunction between the under categories  $E\downarrow\EA$ and $A\downarrow\POG_u$, which we denote for the sake of simplicity $\Gr\dashv \Gamma$ as well. The functor $\Gr$ sends an object $g\colon E\rightarrow F$ of $E\downarrow\EA$ to $A=\mathrm{Gr}(E)\rightarrow \mathrm{Gr}(F)$, while the functor $\Gamma$ sends an object $h\colon (A,u)\rightarrow (B,v)$ of $A\downarrow\POG_u$ to $\bar{h}$ which is a composition $E\rightarrow \Gamma(\mathrm{Gr}(E),u)\rightarrow \Gamma(B,v)$ of the unit $\eta_E$ and $\Gamma(h)$. The adjunction rule is given by a bijective correspondence between the following two triangles, which is induced by the naturality of the original adjunction:
\begin{equation}\label{diag:underAdj}
    \centering
\begin{tikzcd} 
{\Gr(F)} \arrow[rr]&&{(B,v)}&{F} \arrow[rr]&&{\Gamma(B,v)}\\ 
&{\Gr(E)=A}\arrow[lu,"\Gr(g)"]\arrow[ru,"h"]&&&{E}\arrow[lu,"g"]\arrow[ru,"\bar{h}"]&
 \end{tikzcd}
\end{equation}
Let us consider the following square of adjunctions:
\begin{equation}\label{diag:com1}
    \centering
\begin{tikzcd} {E\downarrow\EA} \arrow[dd,bend right=15,"\Gr"'{name=LGr}]\arrow[rr, bend right=15, "{[E,-]_{-}}"'{name=RU}]&&{\EA} \arrow[dd,bend right=15,"\Gr"'{name=RGr}]\arrow[ll, bend right=15, "E\otimes-"{name=LU, above}]\\ 
&&\\
A\downarrow\POG_u\arrow[uu,bend right=15,"\Gamma"'{name=LGa}] \arrow[rr, bend right=15, "{[A,-]_{-}}"'{name=D,below}] &&\POG_u\arrow[ll, bend right=15, "A\otimes-"{name=U, above}]\arrow[uu, bend right=15,"\Gamma"'{name=RGa}] \arrow[from=U, to=D, symbol=\dashv]\arrow[from=LU, to=RU, symbol=\dashv] 
\arrow[from=LGr, to=LGa, symbol=\dashv]\arrow[from=RGr, to=RGa, symbol=\dashv] \end{tikzcd}
\end{equation}
We will prove that the square of adjoints commutes (up to isomorphism). By the uniqueness of a left (right, resp.) adjoint, once we prove the square of the right adjoints or the square of the left adjoints commutes, we yield the commutativity of the other square as well. For the case of left adjoints, the commutativity translates to a desired isomorphism
 \begin{equation}\label{eq:isoGr}
        \Gr(E)\otimes\Gr(F)\cong\Gr(E\otimes F).
    \end{equation}
However, tensor products are rather complicated objects, so we prefer to investigate the square of the right adjoints. In this case, we need to establish an isomorphism between two explicitly defined effect algebras of certain mappings:
\begin{equation}\label{eq:iso}
    [E,\Gamma(B,v)]_{\bar{h}}\cong \Gamma([\Gr(E),B],h).
\end{equation}
\begin{theorem}\label{thm:thereIsIso}
    There exist isomorphisms~\eqref{eq:isoGr} and \eqref{eq:iso}.
\end{theorem}
\begin{proof}
    As we argue above, if we establish the isomorphism~\eqref{eq:iso} natural in $h$, we also obtain an isomorphism~\eqref{eq:isoGr} natural in $F$. The right-hand side of~\eqref{eq:iso}, for a given $h\colon \Gr(E)\rightarrow B$, equals the following set of positive group homomorphisms $\{g\colon \mathrm{Gr}(E)\rightarrow B\mid g\leq h\}$. While the left-hand side equals the following set of generalized effect algebra homomorphisms $\{f\colon E\rightarrow \Gamma(B,v)\mid f\leq \bar{h}\}$.

Hence, the isomorphisms~\eqref{eq:iso} come from the natural isomorphism~\eqref{eq:internalAdj} by cutting at $\bar{h}$ (on the left-hand side) and at $h$ (on the right-hand side):
\begin{equation}\label{eq:adjNat}
 \begin{split}
        \mathrm{Gr}(E)\xrightarrow{g} B, g\leq h\\
        \hline
         E\xrightarrow{\bar{g}}\Gamma(B,v), \bar{g}\leq \bar{h}
    \end{split}
\end{equation}
\end{proof}
Let us present Wehrung's example of two unital Abelian po-groups with interpolation, whose tensor product in $\POG_u$ does not have interpolation.
\begin{example}[see~\cite{W}, Example 1.5.]\label{ex:Werung}
    Define a po-group $G=\{(x,y,z)\in\mathbb{Z}^3\vert x+y+z\equiv 0 \text{(mod. 2)}\}$ with positive cone $G^+=\{0\}\times 2\mathbb Z\times 2\mathbb Z$. Set $A=\mathbb Q\overrightarrow{\times} G$, where $\mathbb Q$ has the usual structure of an Abelian po-group. Next, define po-group $H=\mathbb Z$ with positive cone $H^+=2\mathbb Z$ and set $B=\mathbb Q\overrightarrow{\times} H$. Then $A$, $B$ are interpolation Abelian po-groups with order units $u_A=(1,0,0,0)$ and $u_B=(1,0)$, respectively. However, their tensor product in $\POG_u$ does not satisfy interpolation. 
\end{example}
\begin{corollary}
    In general, the tensor product in $\EA$ does not preserve (RDP).
\end{corollary}
\begin{proof}
Let $(A,u),(B,v)$ be unital Abelian po-groups from Example~\ref{ex:Werung} and denote the corresponding effect algebras $E=\Gamma(A,u)$ and $F=\Gamma(B,v)$. As both $A$ and $B$ satisfy interpolation, we have $A\cong\Gr(E)$ and $B\cong \Gr(F)$ by Theorem~\ref{thm:RDP}. Assume, for the sake of contradiction, that $E\otimes F$ satisfies (RDP); then $\Gr(E\otimes F)$ satisfies interpolation. However, the last is, according to Theorem~\ref{thm:thereIsIso} isomorphic to $\Gr(E)\otimes\Gr(F)\cong A\otimes B$, which is not an interpolation group by the assumption. Thus, we have arrived at a contradiction. 
\end{proof}
\section{Construction of universal group $\Gr$ is strong monoidal}
In this section, we will focus on more subtle features from a category theory perspective. In particular, we will show that we can take an isomorphism~\eqref{eq:isoGr} to be natural in both $E$ and $F$. This is an essential step on the way of establishing $\Gr$ as a strong monoidal functor. So far, we only know the isomorphism~\eqref{eq:al1} exists and is natural in $F$ (as it is induced by a certain natural transformation). However, the naturality in $E$ is not evident, and we do not know how it acts on generating pure tensors.

The following Proposition~\ref{prop:greatDeal} establishes a natural candidate for the isomorphism~\eqref{eq:isoGr}. Before we state the proposition, let us introduce some notation which will improve orientation in the further computations. Given an element $x$ of an effect algebra $E$, we denote its corresponding element of $\Gr(E)$ as $[x]$. Note that according to Lemma~\ref{lem:univ}, the elements of the form $[x]$ generate $\Gr(E)$. Similarly, given an element $a$ of a unital Abelian po-group $(A,u)$, with $0\leq a\leq u$, by $\{a\}$ we indicate it is meant as an element of $\Gamma(A,u)$. Formally $\{[x]\}=\eta_E(x)\in\Gamma(\Gr(E),1)$. (In the rest of the article, there is no occurrence of $``\{"$ and $``\}"$ in the usual set-theoretical meaning, so there is no danger of confusion.) 
\begin{proposition}\label{prop:greatDeal}
 For each pair of effect algebras $E$ and $F$, there is a  homomorphism 
\begin{equation}\label{eq:natDirection}
\gamma_{E,F}\colon\Gr(E\otimes F)\rightarrow \Gr(E)\otimes\Gr(F)
\end{equation}
which acts (on generating elements) as 
\begin{equation}\label{eq:deter}
    [x\otimes y]\mapsto [x]\otimes [y],
\end{equation}
and which is natural in both $E$ and $F$. Moreover, diagrams~\eqref{eq:unitality} commute, where $A=\Gr(E)$, $B=\Gr(F)$, and $\lambda, \rho$ ($\lambda', \rho'$, resp.) come from Definition~\ref{def:monoidalCat} for monoidal category $\EA$ ($\POG_u$, resp.).
\begin{equation}\label{eq:unitality}
    \centering
    \begin{tikzcd}[scale cd=0.98]
        {\Gr(\mathbf{2}\otimes F)}\arrow[r,"\gamma_{\mathbf{2},F}"]&{\Gr(\mathbf{2})\otimes\Gr(F)} &{\Gr(E\otimes\mathbf{2})}\arrow[r,"\gamma_{E,\mathbf{2}}"]&{\Gr(E)\otimes\Gr(\mathbf{2})}\\
        {\Gr(F)}\arrow[u,"\Gr(\lambda^{-1}_F)"]\arrow[r,"(\lambda_B')^{-1}"]&{\mathbb{Z}\otimes \Gr(F)}\arrow[u,"!\otimes 1_B"] &{\Gr(E)}\arrow[u,"\Gr(\rho_E^{-1})"]\arrow[r,"(\rho_A')^{-1}"]&{\Gr(E)\otimes\mathbb{Z}}\arrow[u,"1_A\otimes !"]
    \end{tikzcd}
\end{equation}
 \end{proposition}
 \begin{proof}
     We begin with an obvious bihomomorphism $$\Gr(E)\times\Gr(F)\rightarrow \Gr(E)\otimes\Gr(F)$$ in $\POG_u$. It restricts to a bihomomorphism 
     \begin{align}\label{eq:al1}
\Gamma(\Gr(E),1)\times\Gamma(\Gr(F),1)&\rightarrow \Gamma(\Gr(E)\otimes\Gr(F),1\otimes 1),\\
         (\{[x]\},\{[y]\}) &\mapsto \{[x]\otimes[y]\}\nonumber.
     \end{align}
      Bihomomorphism \eqref{eq:al1} induces an effect algebra homomorphism 
      \begin{align*}
          \Gamma(\Gr(E),1)\otimes\Gamma(\Gr(F),1)&\rightarrow \Gamma(\Gr(E)\otimes\Gr(F),1\otimes 1),\\
          \{[x]\}\otimes\{[y]\} &\mapsto \{[x]\otimes[y]\}.
      \end{align*}
     Then we precompose the last homomorphism with $\eta_E\otimes\eta_F\colon x\otimes y\mapsto \{[x]\}\otimes\{[y]\}$, which yields
           \begin{align*}
          E\otimes F&\rightarrow \Gamma(\Gr(E)\otimes\Gr(F),1\otimes 1),\\
          x\otimes y &\mapsto \{[x]\otimes[y]\}.
      \end{align*}
     Finally, using the adjunction~\eqref{diag:adj1}, we get the desired~\eqref{eq:natDirection}.

    To prove the naturality, we need to verify for each $g\colon E\rightarrow E'$ and $h\colon F\rightarrow F'$
     the equality
     $$\gamma_{E',F'}\circ
     \Gr(g\otimes h)=\Gr(g)\otimes
     \Gr(h)\circ \gamma_{E,F}.$$ 
     According to prescription~\eqref{eq:deter}, the left-hand side acts as $[x\otimes y]\mapsto [g(x)\otimes h(y)]\mapsto [g(x)]\otimes [h(y)]$, while the right-hand side acts as $[x\otimes y]\mapsto [x]\otimes[y]\mapsto [g(x)]\otimes [h(y)]$. Consequently, both sides agree on generating elements and so are equal. 

     One can easily prove the diagrams~\eqref{eq:unitality} commute directly by evaluating them at $[x]\in\Gr(E)$ and $[y]\in\Gr(F)$, respectively. 
 \end{proof}

Let us fix for a while concrete $E\in\EA$ and $A:=\Gr(E)$. Further, let us denote $L_1\colon\EA\rightarrow A\downarrow\POG_u$ the composition of the clockwise path of the left adjoints in~\eqref{diag:underAdj} and $L_2\colon\EA\rightarrow A\downarrow\POG_u$ 
the composition of the counterclockwise path of the left adjoints in~\eqref{diag:underAdj}
\begin{align}
    L_1&\colon F\mapsto (\Gr(\iota_{E,F})\colon A\rightarrow \Gr(E\otimes F)),\label{eq:L1} \\ 
    L_2&\colon F\mapsto (\iota'_{\Gr(E),\Gr(F)}\colon A\rightarrow \Gr(E)\otimes\Gr(F)).\label{eq:L2}
\end{align}
Similarly, we denote $R_1$ and $R_2$ the compositions of the right adjoints, so $L_i\dashv R_i$, for $i=1,2$.

Next, consider the following commutative diagram 
\begin{equation}\label{diag:cos}
    \centering
\begin{tikzcd} 
{\Gr(E\otimes F)}\arrow[r,"\gamma_{E,F}"]&{\Gr(E)\otimes \Gr(F)}\\ 
{\Gr(E\otimes\mathbf{2})}\arrow[u]\arrow[r,"\gamma_{E,\mathbf{2}}"]&{\Gr(E)\otimes\Gr(\mathbf{2})}\arrow[u]\\
{\Gr(E)}\arrow[u,"\Gr(\rho^{-1}_E)"]\arrow[r,"{(\rho'_{A})}^{-1}"]&{\Gr(E)\otimes \mathbb{Z}}\arrow[u,"1\otimes !"]
 \end{tikzcd}
 \end{equation}
where the upper square commutes by naturality of $\gamma$ and the lower square commutes by~\eqref{eq:unitality}. The left leg of~\eqref{diag:cos} equals $$\Gr(\iota_{E,F})=
L_1(F)$$ defined in~\eqref{eq:L1}. Whereas the composition of the lower-right path equals 
$$\iota'_{\Gr(E),\Gr(F)}=L_2(F).$$
Consequently, commutative diagram~\eqref{diag:cos} implies that $\gamma_{E,(-)}$ defines a natural transformation \begin{equation}\label{eq:defGammaE}
    \gamma_{E,(-)}\colon L_1\Rightarrow L_2.
\end{equation} 

    Recall we have established an isomorphism~\eqref{eq:isoGr} by the argument about the uniqueness of a left adjoint. We would like the isomorphism to coincide with (the inverse of) the natural homomorphism from Proposition~\ref{prop:greatDeal}. To achieve this, we refine the argument about the uniqueness of a left adjoint using the concept of conjugate natural transformations, which is introduced in the following theorem (for the proof see section  I.7. in~\cite{La}).
\begin{theorem}[\cite{La}]\label{thm:conjugation}
    Let $L,L'\colon\mathcal{C}\rightarrow\mathcal{D}$ be two functors having right adjoints $R,R'$ (respectively) and $\alpha\colon L\Rightarrow L'$ be a natural transformation. Then there exists a unique natural transformation $\beta\colon R'\rightarrow R$, so that for each $X\in\mathcal{C}$ and $Y\in\mathcal{D}$, the following diagram commutes:
    \begin{equation}\label{diag:conjungation}
    \centering
\begin{tikzcd} 
{\mathrm{Hom(L'(X),Y)}}\arrow[r,leftrightarrow,"\cong"]\arrow[d,"{\mathrm{Hom}(\alpha_X,Y)}"]&{\mathrm{Hom}(X,R'(Y))}\arrow[d,"{\mathrm{Hom}(X,\beta_Y)}"]\\ 
{\mathrm{Hom(L(X),Y)}}\arrow[r,leftrightarrow,"\cong"]&{\mathrm{Hom}(X,R(Y))}
 \end{tikzcd}
 \end{equation}
 We call such $\alpha$ and $\beta$ conjugate. The relation of conjugation provides a bijection between $\mathrm{Nat}(L,L')$ and $\mathrm{Nat}(R',R)$ which restricts to a bijection between natural isomorphisms. 
 \end{theorem}
Note that the last theorem gives us a strong control over natural transformations of the left adjoints $L_1, L_2$ via investigating natural transformations of the right adjoints $R_1, R_2$, which are in our case easier to work with.  
 
 There are several criteria that describe when two natural transformations are conjugate. We will apply the following one.
 \begin{theorem}[\cite{La}, Thm. 2 (sec. I.7.)]\label{thm:criOfCon}
     Let $L,L'\colon\mathcal{C}\rightarrow\mathcal{D}$ be two functors having right adjoints $R,R'$ (respectively) and denote $\eta$ and $\eta'$ the corresponding units of the two adjunctions. Two natural transformations $\alpha\colon L\Rightarrow L'$ and $\beta\colon R'\Rightarrow R$ are conjugate if and only if the following square of natural transformations commutes:
       \begin{equation}\label{diag:criOfCon}
    \centering
\begin{tikzcd} 
{\mathbf{1}_\mathcal{C}}\arrow[r,"\eta"]\arrow[d,"\eta'"]&{RL}\arrow[d,"R\alpha"]\\ 
{R'L'}\arrow[r,"\beta L'"]&{RL'}
 \end{tikzcd}
 \end{equation}
 \end{theorem}

 The following proposition establishes a great deal of the result that $\Gr$ is a monoidal functor.

\begin{proposition}\label{prop:isConv}
        Natural transformation $\gamma_{E}\colon L_1\Rightarrow L_2$, given by~\eqref{eq:defGammaE}, is conjugate to a natural isomorphism $\beta_E\colon R_2\Rightarrow R_1$ which is given by the isomorphism~\eqref{eq:iso}, i.e., for $h\colon A=\Gr(E)\rightarrow (B,v)$ an object of $A\downarrow\POG_u$, there is
        \begin{align}
            \beta_{E,h\colon A\rightarrow B}\colon \Gamma([A,B],h)&\rightarrow [E,\Gamma(B,v)]_{\bar{h}},\\
            \{g\colon [x]\mapsto g([x])\}&\mapsto \bar{g}\colon x\mapsto \{g([x])\}.\label{eq:pres3}
        \end{align}
\end{proposition}
\begin{proof}
According to Proposition \ref{thm:criOfCon}, it is enough to verify that for any effect algebra $F$, the square~\eqref{diag:criOfC} commutes.
 \begin{equation}\label{diag:criOfC}
    \centering
\begin{tikzcd}
{F}\arrow[r,"\eta_F"]\arrow[d,"\eta'_F"]&{R_1L_1(F)}\arrow[d,"R_1(\gamma_{E,F})"]\\ 
{R_2L_2(F)}\arrow[r,"\beta_{E,L_2(F)}"]&{R_1L_2(F)}.
 \end{tikzcd}
 \end{equation}
The unit $\eta_F$ is by definition as follows:
\begin{align*}
    \eta_F\colon F&\rightarrow [E,\Gamma(\Gr(E\otimes F),[1\otimes 1])]_{f_0},\\
    y&\mapsto (x\mapsto\{[x\otimes y]\}),
\end{align*}
where $f_0\colon x\mapsto \{[x \otimes 1]\}$. If we compose it with $R_1(\gamma_{E,F})$, we get the homomorphism given by the prescription 
\begin{align}
    R_1(\gamma_{E,F})\circ \eta_F\colon F&\rightarrow [E,\Gamma(\Gr(E)\otimes \Gr(F),[1]\otimes [1])]_f,\label{eq:pres1}\\
    y&\mapsto (x\mapsto\{[x]\otimes [y]\}), \notag
\end{align}
where $f\colon x\mapsto \{[x]\otimes[1]\}$. The other way around, $\eta'_F$ is by definition as follows: 
\begin{align*}
   \eta'_F\colon F&\rightarrow \Gamma([A,\Gr(E)\otimes\Gr(F)],h),\\
    y&\mapsto \{[x]\mapsto [x]\otimes [y]\},
\end{align*}
where $h\colon [x]\mapsto [x]\otimes [1]$. 
Under composition with $\beta_{E,L_2(F)}$ (we apply~\eqref{eq:pres3} to $g\colon [x]\mapsto [x]\otimes [y]$), we get
\begin{align}
   \beta_{E,L_2(F)}\circ \eta'_F\colon F&\rightarrow [E,\Gamma(\Gr(E)\otimes \Gr(F),[1]\otimes [1])]_{\overline{h}},\label{eq:pres2}\\
    y&\mapsto (x\mapsto\{[x]\otimes [y]\}), \notag
\end{align}
where $\overline{h}\colon x\mapsto \{[x]\otimes[1]\}$.
The two compositions lead to the same prescriptions~\eqref{eq:pres1} and~\eqref{eq:pres2}, therefore the square~\eqref{diag:criOfC} in concern commutes.
\end{proof}

\begin{theorem}\label{def:grIsMonoidal}
    The functor $\Gr\colon \EA\rightarrow \POG_u$ is a strong monoidal functor. Here $\epsilon$ is the unique homomorphism $\epsilon:\mathbb{Z}\rightarrow \Gr(\mathbf{2})$ and $\mu$ has components $\mu_{E,F}$, $E,F\in \EA$, given by the prescription 
    \begin{equation}\label{eq:mu}
        \mu_{E,F}\colon[x]\otimes[y]\mapsto [x\otimes y].
    \end{equation} 
\end{theorem}
\begin{proof}
By Proposition~\ref{prop:isConv}, for each effect algebra $E$, natural transformation $\gamma_{E}$ and natural isomorphism $\beta_E$ from Proposition~\ref{prop:isConv} are conjugate. Consequently, by Theorem~\ref{thm:conjugation}, $\gamma_E$ is a natural isomorphism as well. It follows, for each $E,F\in\EA$, $\gamma_{E,F}$ has an inverse $\mu_{E,F}:=\gamma_{E,F}^{-1}$, which defines a natural transformation $\mu$ with prescription~\eqref{eq:mu}.  

We have to verify commutative diagrams from Definition~\ref{defn:monoidalFunctor}. The two square diagrams are already established in Proposition~\ref{prop:greatDeal}, one only needs to invert some isomorphisms. For the hexagon diagram, it is trivial to verify it holds on elements of the form $([x]\otimes [y])\otimes[z]$. As these elements generate the domain po-group, the hexagon commutes.
\end{proof}
\section{Further work}
The technique introduced in this article could be applied to analyze other adjunctions between categories, similar to the approach in Section~\ref{sec:4} for the categories 
 $\EA$ and $\POG_u$. In particular, an adjunction between $\EA$ and some convenient category of partial monoids, or between $\EA$ and the category of monotone $\sigma$-complete effect algebras studied in~\cite{Lach}. 
 
 Another important direction for further research is providing a concrete method for computing the tensor product of two effect algebras. While this may be challenging for a general pair of effect algebras, specific cases appear more tractable. For instance, the computation of the tensor product of the real unit interval with itself $[0,1]_\mathbb{R}\otimes[0,1]_\mathbb{R}$ is a promising starting point for such an investigation.



\subsection*{Data availability}
Data sharing is not applicable to this article as datasets were neither generated nor analysed.

\medskip
\subsection*{Compliance with ethical standards}
The authors declare that they have no conflict of interest.


\bibliographystyle{spmpsci}
\bibliography{bibliography}
\end{document}